\documentclass[11pt]{article}
\usepackage{graphicx}
\usepackage{dcolumn}
\usepackage{amsmath}
\usepackage{epsfig}

\input pubboard/babarsym
\input Definitions

\newcommand{\BABARPubYear}    {04}

\newcommand{\BABARConfNumber} {19}
\newcommand{\SLACPubNumber} {10640}

\long\def\inst#1{\par\nobreak\kern 4pt\nobreak
    {\it #1}\par\vskip 10pt plus 3pt minus 3pt}

\begin{document}
{\pagestyle{empty}

\begin{flushright}
\babar-CONF-\BABARPubYear/\BABARConfNumber \\
SLAC-PUB-\SLACPubNumber \\
July 2004 \\
\end{flushright}

\par\vskip 5cm

\begin{center}
\Large \bf A Measurement of {\em CP} Violating Asymmetries 
in \boldmath$B^0\to\ffzeroKs$ Decays
\end{center}
\bigskip

\begin{center}
\large The \babar\ Collaboration\\
\mbox{ }\\
\today
\end{center}
\bigskip \bigskip

\centerline{\small{\bf Abstract}} 
\vspace{0.1cm}
\noindent
{\normalsize
We present preliminary measurements of the 
\CP-violating asymmetries in the decay $B^0\to\fzero(\to\pi^+\pi^{-})\KS$.
The results are obtained from a data sample of $209\times10^6$ 
$\FourS \to B\Bbar$ decays collected with the \babar\  detector 
at the \pep2 asymmetric-energy $B$~Factory at SLAC.
From a time-dependent maximum-likelihood fit we measure 
the mixing-induced \CP\ violation parameter 
$S=-0.95^{+0.32}_{-0.23} \pm 0.10$
and the direct \CP\ violation parameter 
$C=-0.24\pm 0.31         \pm 0.15 $, 
where the first errors are statistical and the second systematic.
}

\vfill
\begin{center}

Submitted to the 32$^{\rm nd}$ International Conference on High-Energy Physics, ICHEP 04,\\
16 August---22 August 2004, Beijing, China

\end{center}

\vspace{1.0cm}
\begin{center}
{\em Stanford Linear Accelerator Center, Stanford University, 
Stanford, CA 94309} \\ \vspace{0.1cm}\hrule\vspace{0.1cm}
Work supported in part by Department of Energy contract DE-AC03-76SF00515.
\end{center}

\newpage
} 

\begin{center}
\small

The \babar\ Collaboration,
\bigskip

%
B.~Aubert,
R.~Barate,
D.~Boutigny,
F.~Couderc,
J.-M.~Gaillard,
A.~Hicheur,
Y.~Karyotakis,
J.~P.~Lees,
V.~Tisserand,
A.~Zghiche
\inst{Laboratoire de Physique des Particules, F-74941 Annecy-le-Vieux, France }
% ...

\end{center}\newpage


\section{Introduction}
\label{sec:Introduction}
In the Standard Model (SM), \CP violation arises from a single
phase in the three-generation Cabibbo-Kobayashi-Maskawa 
quark-mixing matrix~\cite{CKM}.  Possible indications of physics
beyond the SM may be observed in time-dependent 
\CP asymmetries of  $B$ decays dominated by penguin-type diagrams
to states such as $\phi K^0$,
$\eta^\prime K^0$, $K^+K^-K^0$, and $\fzero K^0$ \cite{NewPhys}.
Neglecting CKM-suppressed amplitudes,
these decays carry  the same weak phase as the decay 
$\Bz\to\jpsi K^0$ \cite{phases}. As a consequence, their mixing-induced 
\CP-violation parameter is expected to be 
$-\eta_f\times\stwob = -\eta_f\times0.74\pm0.05$ \cite{HFAG} 
in the SM, where 
$\beta \equiv \arg \left[\, -V_{\rm cd}^{}V_{\rm cb}^* / V_{\rm td}^{}V_{\rm tb}^*\, \right]$
and $\eta_f$ is the \CP eigenvalue of the final state 
$f$, which is +1 for $\ffzeroKs$.  
There is no direct \CP violation expected
in these decays since they are dominated by a single amplitude in the SM.
Due to the large virtual masses occurring in the penguin loops,
additional diagrams with non-SM heavy particles in the loops and new
\CP-violating phases may contribute. Measurements 
of \CP violation in these channels and their comparisons with the SM 
expectation are therefore sensitive probes for physics beyond the SM.

\renewcommand{\thefootnote}{\fnsymbol{footnote}}

We present the preliminary results of an update of a measurement~\cite{f0ksprl}
of \CP-violating asymmetries in the 
penguin-dominated decay $B^0\rar\fzeros\KS$ \footnote[2]{
Throughout the paper $\fzeros$ refer to the $\fzero$ and its decay to $\pi^+\pi^-$.
In addition, charge conjugate decay modes are 
assumed unless explicity stated.} from a time-dependent
maximum-likelihood analysis. We restrict the analysis to the region 
of the $\pi^+ \pi^- \KS$ Dalitz plot that is dominated by the $\fzeros$ and
we refer to this as the quasi-two-body (Q2B) approach.  Effects due to the interference 
between the $\fzeros$ and the other resonances in the Dalitz plot are taken as 
systematic uncertainties.  

\begin{figure}[b]
\begin{center}
\includegraphics[width=10cm]{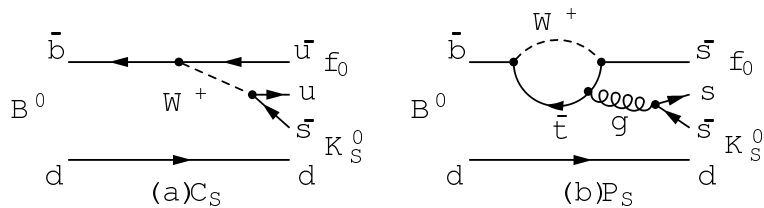}
\caption{\em \label{fig:feyn}
	The color-suppressed tree (a) and dominant gluonic penguin (b) 
  are diagrams that could contribute to the decay \fzeroKs.}
\end{center}
\end{figure}

The structure of the scalar meson \fzeros\ has been discussed
for decades and is still obscure. There were attempts to
interpret it as $K\Kbar$ molecular states~\cite{kkbar}, 
four-quark states~\cite{4quark} and normal \qqbar states~\cite{qqbar}. 
However, recent studies of 
$\phi\to\gamma\fzeros\ (\fzeros\to\gamma\gamma)$~\cite{phisangle1a,phisangle1b}
and $D_s^+ \to\fzeros\pip$~\cite{phisangle2} decays favor the \qqbar 
state models. In this interpretation the flavor content of the $\fzeros$ is
given by $\fzeros=\cos(\phi_s) s\bar s+\sin(\phi_s) n\bar n$, with
$n\bar n=(u\bar u+d\bar d)/\sqrt{2}$. A mixing phase of
$\phi_s =-48^\circ\pm6^\circ$ has been experimentally determined from 
$\phi\to\gamma\fzeros$ decays~\cite{phisangle1b}.  If the assumption 
is true that the $\fzeros$ state has a sizable content of $s\bar s$, 
then the decay $\Bz \to \fzeros\KS$ would be dominated by the penguin transition, 
$b\to s\overline{s}s$ (\cf\  Fig.~\ref{fig:feyn}(b)). Thus, we expect that a 
measurement of mixing-induced CP violation leads to $\Sfzks\simeq-\stwob$,
where $\Sfzks$ is the coefficient of the sine modulation term\cite{NewPhys}.

The data  used in this analysis were accumulated with the \babar\ 
detector~\cite{bib:babarNim} at the \pep2 asymmetric-energy $e^+e^-$ 
storage ring at SLAC. The data sample consists of  an integrated 
luminosity of $192\invfb$  collected at the \FourS resonance
 (``on-resonance'') corresponding to $209\times10^{6}$ 
$\B\Bbar$ pairs,
and $11.8\invfb$ collected about $40\mev$ below 
the~\FourS (``off-resonance''). In Ref.~\cite{bib:babarNim} we describe 
the silicon vertex tracker and drift chamber used for track and vertex
reconstruction, and the Cherenkov detector (DIRC), the electromagnetic 
calorimeter (EMC), and the instrumented flux return (IFR) used for particle 
identification.

If we denote by  $\deltat$  the difference between the proper 
times of the decay of  the fully reconstructed $B^0\rar f_0\KS$ ($\Bz_{\rm rec}$)
and the decay  of the  other meson ($\Bz_{\rm tag}$)\footnote
{ The $\Bz_{\rm tag}$ is so called because its flavor is determined using the
tagging algorithm of Ref.~\cite{bib:BabarS2b}.
},  
the time-dependent decay rate $f_{{\rm Q_{tag}}}$ is given by 
\beqn
\label{eq:theTime}
f_{{\rm Q_{tag}}}(\deltat)  =  
        \frac{e^{-\left|\deltat\right|/\tau}}{4\tau}
        \bigg( 1 + {\rm Q_{tag}} S\sin(\deltamd\deltat) -\;{\rm Q_{tag}} C\cos(\deltamd\deltat)
        \bigg)
\eeqn
where   $Q_{\rm tag}= 1(-1)$ when the tagging meson $\Bz_{\rm tag}$
is a $\Bz(\Bzb)$, $\tau$ is the mean \Bz lifetime, and $\deltamd$ is 
the $\BzBzb$ oscillation frequency corresponding to the mass difference. 
The parameter $S$ is non-zero if there is  mixing-induced \CP violation, 
while a non-zero value for $C$ would indicate direct \CP violation.

\section{Candidate Selection}
\label{sec:selection}
We reconstruct $B^0\rar f_0 (\rar\pi^+\pi^-) \KS$ candidates from combinations 
of two  tracks and a $\KS$~decaying to $\pi^+\pi^-$.
For the $\pi^+\pi^-$ pair  from the $\fzeros$ candidate, 
we use information  from the tracking system, EMC, and DIRC to 
remove tracks consistent with electron, kaon, and proton hypotheses.  
In addition, we require at least one track to have a          
signature in the IFR that is inconsistent with the muon hypothesis.
The mass of the 
$\fzeros$ candidate must satisfy $0.86<m(\pi^+\pi^-)<1.10\gevcc$.  To reduce
combinatorial background from low energy pions, we 
require $|\cos{\theta(\pi^+)}|<0.9$, where  $\theta(\pi^+)$ 
is the angle between the positive pion direction in the $\fzeros$  rest frame
 and the $\fzeros$ flight direction in the laboratory frame.  
The $\KS$ candidate is required to have a mass within $10\mevcc$ of 
the nominal $K^0$ mass \cite{PDG2002} and a decay vertex separated from the $B^0$ decay vertex
 by at least five standard deviations. In addition, the cosine 
of the angle between the $\KS$ flight direction and the vector between 
the $\fzeros$ and the $\KS$ vertices must be greater than 0.99.

Two kinematic variables are used to discriminate between signal-$B$ 
decays and combinatorial background. One variable is the difference 
$\de$ between the measured center-of-mass (CM) energy of the $B$~candidate and $\sqrt{s}/2$, where $\sqrt{s}$ is the CM energy. The other variable is the 
beam-energy-substituted mass 
$\mes\equiv\sqrt{(s/2+{\mathbf {p}}_i\cdot{\mathbf{p}}_B)^2/E_i^2-{\mathbf {p}}_B^2},$
where the $B$ momentum ${\mathbf {p}}_B$ and the four-momentum of the 
initial state ($E_i$, ${\mathbf {p}}_i$) are defined in the laboratory 
frame. We require $5.23 < \mes <5.29\gevcc$ and $|\de|<0.1\gev$.

Continuum $e^+e^-\to q\bar{q}$ ($q = u,d,s,c$) events are the dominant 
background.  To enhance discrimination between signal and continuum, we 
use a neural network (NN) to combine four  variables: the 
cosine of the angle between the $B^0_{\rm rec}$ direction 
and the beam axis in the CM,
the cosine of the angle between the thrust axis of the
$B^0_{\rm rec}$  candidate
and the beam axis, and the zeroth and second angular moments $L_{0,2}$
of the energy flow about the $B^0_{\rm rec}$ thrust axis.  The moments
are defined by $L_j=\sum_i p_i \times |\cos{\theta_i}|^j$, where $\theta_i$
is the angle with respect to the $B^0_{\rm rec}$ thrust axis of the track
or neutral cluster $i$, and $p_i$ is its momentum.  The sum excludes the 
$B^0_{\rm rec}$ candidate. The thrust axis
is defined as the direction that maximizes the sum of the longitudinal
momenta of the $B^0_{\rm rec}$ daughters.
The NN is trained 
with off-resonance data and
simulated signal events. The final sample of signal candidates 
is selected with a cut on the NN output $>-1.5$, which retains 
$\sim 97\%$ and $52\%$ of the signal and continuum, respectively.

The signal efficiency determined from Monte Carlo (MC) 
simulation is $(38.7\pm0.4)\%$.  MC simulation shows that  $4.7\%$ of the 
selected signal events are mis-reconstructed,  
 mostly due to combinatorial
background from low-momentum tracks used to form the $\fzeros$ candidate.
In total, $12586$ on-resonance data events pass all selection criteria.

\section{Background from other \boldmath{\B} Decays}
\label{sec:BBackground}

We use  MC-simulated events to study the background from other $B$
decays. The charmless decay modes are grouped into eight classes with similar 
kinematic and topological properties. The modes that decay to the $\pi^+\pi^-\KS$ 
final state are of particular importance since they have signal-like 
$\de$ and $\mes$ distributions and their decay amplitudes interfere with 
the $\fzeros\KS$ decay amplitude.  Among these modes are $\rho^0\KS$, 
$\fzerop\KS$, $\ftwo\KS$, $K^{*+}\pi^-$ (including other kaon 
resonances decaying to $\KS\pi^+$), and non-resonant $\pi^+\pi^-\KS$ 
decays. The mode $\rho^0\KS$ is particularly important because it has
$\eta_f=-1$ and thus any $\rho^0\KS$ events misidentified as signal will
dilute the observed CP asymmetry in our data.
The inclusive charmless $\pi^+\pi^-\KS$ branching fraction 
$(23.4\pm3.3)\tmsix$~\cite{HFAG}, together with the available 
exclusive measurements~\cite{HFAG},  are used to infer upper limits on the
branching fractions of these decays.  Along with selection efficiencies
obtained from MC, these branching fractions are used to estimate the expected
background.
 The charmed decays  
$B^0\rar D^-\pi^+\rar\KS\pi^-\pi^+$ and $\B^+\rar\Dzb\pi^+\rar\KS\pi^0\pi^+$  
contribute significantly to the selected data sample.  Each of these modes 
is  treated as a separate class. Two additional classes account for 
the remaining neutral and charged $b \to c$ decays.
In the selected data sample we expect $106\pm23$ charmless and $218\pm93$ 
$b \to c$ events.

\section{Maximum-Likelihood Fit}
\label{sec:themlfit}
The time difference $\deltat$ is obtained from the measured distance between 
the $z$ positions (along the beam direction) of the $\Bz_{\rm rec}$ and 
$\Bz_{\rm tag}$ decay vertices, and the boost $\beta\gamma=0.56$ of 
the \epem\  system~\cite{bib:BabarS2b,bib:BabarSin2alpha}. 
To determine the flavor of the $\Bz_{\rm tag}$ 
we use the tagging algorithm of Ref.~\cite{bib:BabarS2b}.
This produces four mutually exclusive tagging categories. We also 
retain untagged events in a fifth category to improve the efficiency 
of the signal selection.

We use an unbinned extended maximum-likelihood fit to extract
the $\fzeros\KS$ event yield, the \CP parameters defined 
in Eq.~(\ref{eq:theTime}), and the $\fzeros$ resonance parameters. 
The likelihood function for the $N_\cat$ candidates tagged in category $k$ is
\begin{equation}
\label{eq:pdfsum}
{\cal L}_k = e^{-N^{\prime}_\cat}\!\prod_{i=1}^{N_\cat}
		\bigg\{ N_{S}\epsilon_\cat\left[
                                (1-f^\cat_\textrm{MR}){\cal P}_{i,\cat}^{S-\textrm{CR}} +
                                f^\cat_\textrm{MR}{\cal P}_{i,\cat}^{S-\textrm{MR}}
                              \right]
	+ N_{C,\cat} {\cal P}_{i,\cat}^{C} 
 	+ \sum_{j=1}^{n_B} N_{B,j} \epsilon_{j,\cat}{\cal P}^{\B}_{ij, \cat}
	\bigg\}
\end{equation}
where $N^{\prime}_\cat$ is the sum of the signal, continuum 
and the $n_B$ $B$-background yields tagged in category $\cat$,
 $N_S$ is the number of 
$\fzeros\KS$ signal events in the sample, $\epsilon_\cat$ is the 
fraction of signal events tagged in category $\cat$, $f^\cat_\textrm{MR}$
is the fraction of mis-reconstructed signal events in tagging category
$\cat$, $N_{C,\cat}$ 
is the number of continuum background events  that are tagged in 
category~$\cat$, and $N_{B,j}\epsilon_{j,\cat}$ is the number of
$B$-background events of class $j$ (see section \ref{sec:BBackground}) 
that are tagged in category~$\cat$.
The \B-background event yields are fixed parameters, with the exception 
of the $D^-\pi^+$ yield.  Since $\Bz\rar D^-\pi^+$ events have a 
characteristic distribution in $\cos{\theta(\pi^+)}$, well 
separated from continuum and $\fzeros\KS$ events, the $D^-\pi^+$ is
 free to vary in the fit along with the signal and continuum yields.
The total likelihood 
${\cal L}$ is the product of the likelihoods for each tagging category.

The probability density functions (PDFs) ${\cal P}_{\cat}^{S-\textrm{CR}}$,  
${\cal P}_{\cat}^{S-\textrm{MR}}$,
${\cal P}_{\cat}^{C}$ and ${\cal P}^{\B}_{j, \cat}$, for correctly reconstructed
signal, mis-reconstructed signal, 
continuum background and $B$-background class $j$, respectively,
are the products of the PDFs of six discriminating variables.
The correctly reconstructed signal PDF is thus given by:
$  {\cal P}_\cat^{S-\textrm{CR}} = 
	{\cal P}^{S-\textrm{CR}}(\mes)\cdot 
	{\cal P}^{S-\textrm{CR}}(\de) \cdot
 	{\cal P}_\cat^{S-\textrm{CR}}(\NN) \cdot 
	{\cal P}^{S-\textrm{CR}}(|\cos{\theta(\pi^+)|}) \cdot 
	{\cal P}^{S-\textrm{CR}}(m(\pi^+\pi^-)) 
	\cdot {\cal P}_\cat^{S-\textrm{CR}}(\deltat)
$,
where ${\cal P}_\cat^{S-\textrm{CR}}(\deltat)$ contains the time-dependent 
\CP parameters defined in Eq.~(\ref{eq:theTime}), diluted by the 
effects of mis-tagging and the~$\deltat$ resolution. 

The fractions of mis-reconstructed signal events in each tagging 
category are estimated by MC simulation.  The $\mes$, $\de$, 
\NN, $|\cos{\theta(\pi^+)}|$, and $m(\pi^+\pi^-)$  PDFs 
for signal and $B$ background are taken from the simulation except for 
the means of the signal Gaussian PDFs for $\mes$ and $\de$ as well as the 
mass and width of the $\fzeros$, which are free to vary in the fit.  We use 
a relativistic Breit-Wigner function to parameterize the $\fzeros$
resonance.  
The $\deltat$-resolution function for signal and $B$-background 
events is a sum of three Gaussian distributions, with parameters 
determined by a fit to fully reconstructed $\Bz$ decays~\cite{bib:BabarS2b}.
The continuum $\deltat$ distribution is parameterized as the sum of 
three Gaussian distributions  with two distinct means and three distinct 
widths,  which are scaled by the $\dt$ per-event error. 
For the $B$-background  modes that are \CP eigenstates, the parameters 
$C$ and $S$ are fixed to 0 and $\pm \sin{2\beta}$, respectively, 
depending on their \CP eigenvalues. For continuum, four tag asymmetries 
and the five yields $N_{C,\cat}$  are free. 
The signal yield, $S$, $C$, and the $\fzeros$ mass and width are 
among the 41 parameters that are free to vary in the fit.  The majority of the 
free parameters are used to describe the shape of the continuum background.

\section{Systematic Errors}
\label{sec:systematics}

\begin{table}[b]
\begin{center}
\caption{ \label{tab:systematics}
        Summary of systematic uncertainties.}
\small
\setlength{\tabcolsep}{0.95pc}
\begin{tabular}{lccc} \hline
&&& \\[-0.3cm]
Error Source &  $\Sfzks$ &  $\Cfzks$\\
\hline
&&& \\[-0.3cm]
Fitting Procedure           & 0.06 ~~  & 0.10 ~~\\
$B$-background              & 0.04 ~~  & 0.08 ~~\\
$\dt$ Model                 & 0.01 ~~  & 0.01 ~~\\
Tagging                     & 0.02 ~~  & 0.01 ~~\\
Signal Model                & 0.02 ~~  & 0.02 ~~\\
DCS Decays                  & 0.01 ~~  & 0.04 ~~\\
$\dmd$ and $\tau$           & 0.00 ~~  & 0.01 ~~\\
Q2B Approximation           & 0.04 ~~  & 0.07 ~~\\
\hline
Sub-total                   & 0.10 ~~  & 0.15 ~~\\

\hline 
\end{tabular}
\end{center}
\vspace{-0.25in}
\end{table}
The contributions to the systematic error on the signal parameters are 
summarized in Table~\ref{tab:systematics}.
To estimate the errors due to the fit procedure, we perform fits on a large 
number of MC samples with  the proportions
of  signal, continuum and $B$-background events measured from data. 
Biases of a few percent
observed in these fits are due to imperfections in the likelihood model 
such as neglected correlations between the discriminating variables of the 
signal and $B$-background PDFs and are assigned as a systematic uncertainty 
of the fit procedure.  The error due to the fit procedure includes these biases
added in quadrature with their statistical errors.
The expected event yields from the $B$-background modes are varied according 
to the uncertainties in the measured or estimated branching fractions.  
Since $B$-background modes may exhibit  \CP violation, the corresponding 
CP parameters are varied within their physical ranges. 
We vary the parameters of the $\dt$ model and tagging fractions
incoherently within their errors and assign the observed changes, summed in quadrature, as a systematic error.
The uncertainties due to the simulated signal PDFs are obtained from a control 
sample of fully reconstructed $B^{0} \rightarrow D^{-}(\to\KS\pi^-) \pi^{+}$ decays. 
The systematic errors  due to interference between the 
doubly-Cabibbo-suppressed  
(DCS) $\bar b \to \bar u c \bar d$ amplitude with the Cabibbo-favored 
$\bar b \to \bar c u \bar d$ amplitude for tag-side $B$ decays have been 
estimated from simulation by varying freely all relevant strong 
phases~\cite{bib:DCSD2003}.  The errors associated 
with $\dmd$ and $\tau$ are estimated by varying these parameters within
the errors on the world average~\cite{PDG2002}.

The systematic error introduced in the Q2B approximation
by ignoring interference effects between the $\fzeros$ and the other 
resonances  in the Dalitz plot is estimated 
from simulation by varying freely all relative strong phases and taking 
the largest observed change in each parameter as the error.  
Eleven resonances are used in this study including 
the three lowest lying $\rho$ resonances, $\fzero$, 
$f_0(1370)$, $f_2(1270)$, 
and the $K^{*\pm}$ and higher kaon states. 
In addition, a non-resonant component is allowed.  
The proportion of each contribution is estimated 
using known exclusive measurements and the inclusive 
$\pi^+\pi^-\KS$ rate.  
The systematic effects due to interference  
are small compared with the  statistical error for $S$ and $C$.


\begin{figure}[p]
  \centerline{ 	\epsfysize5.0cm\epsffile{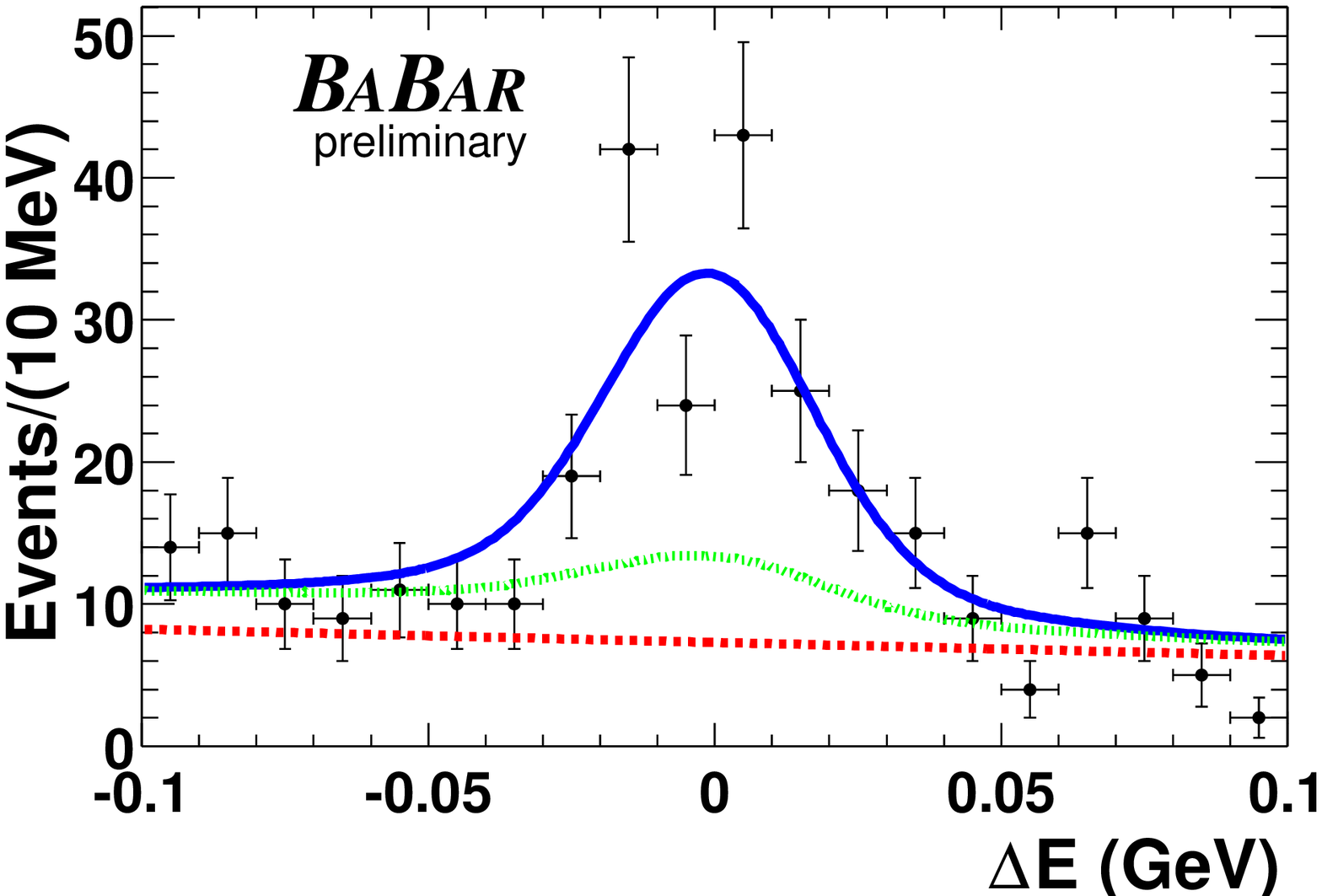}
		\epsfysize5.0cm\epsffile{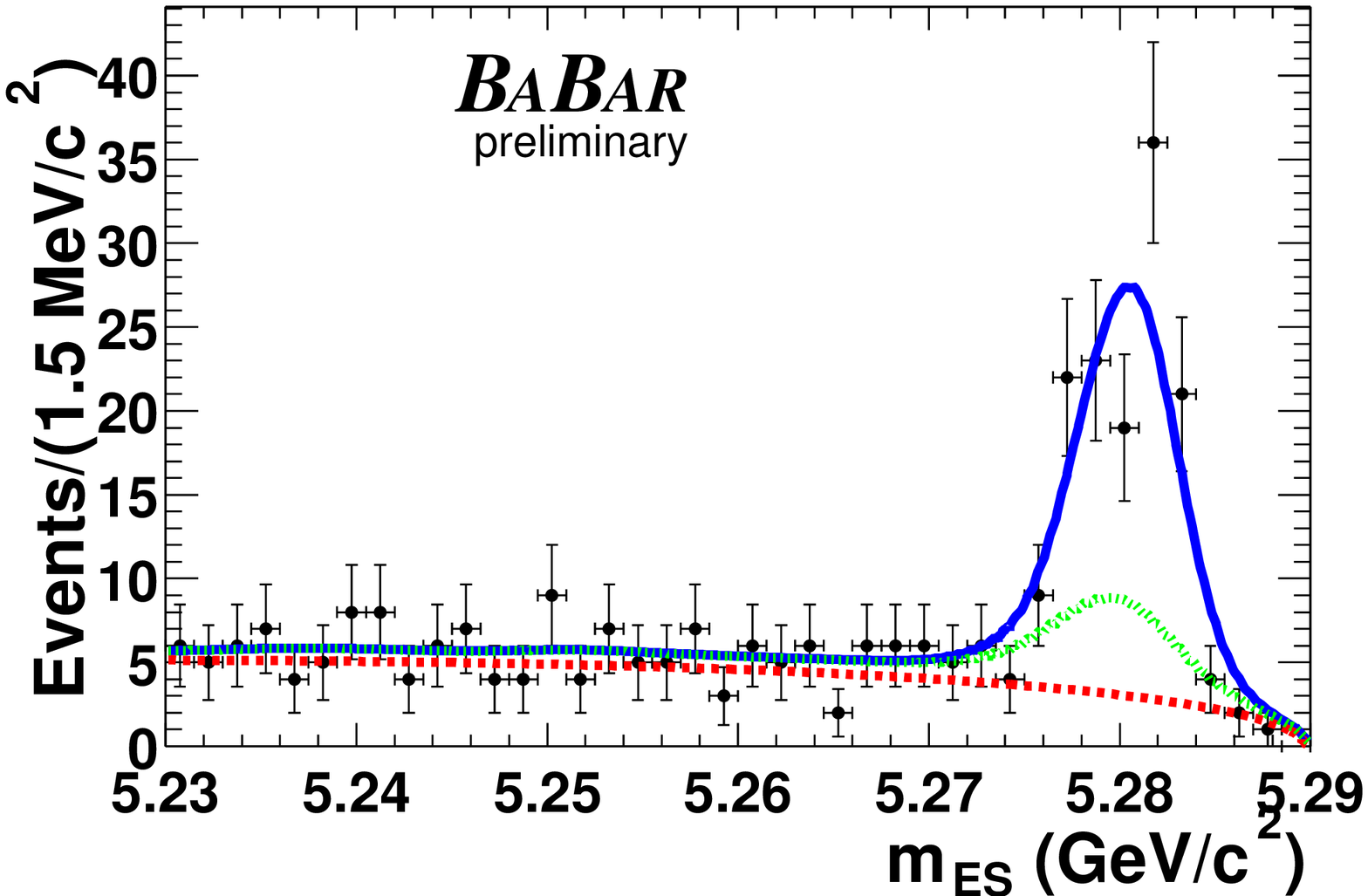}}
\vspace{0.45cm}
  \centerline{ 	\epsfysize5.0cm\epsffile{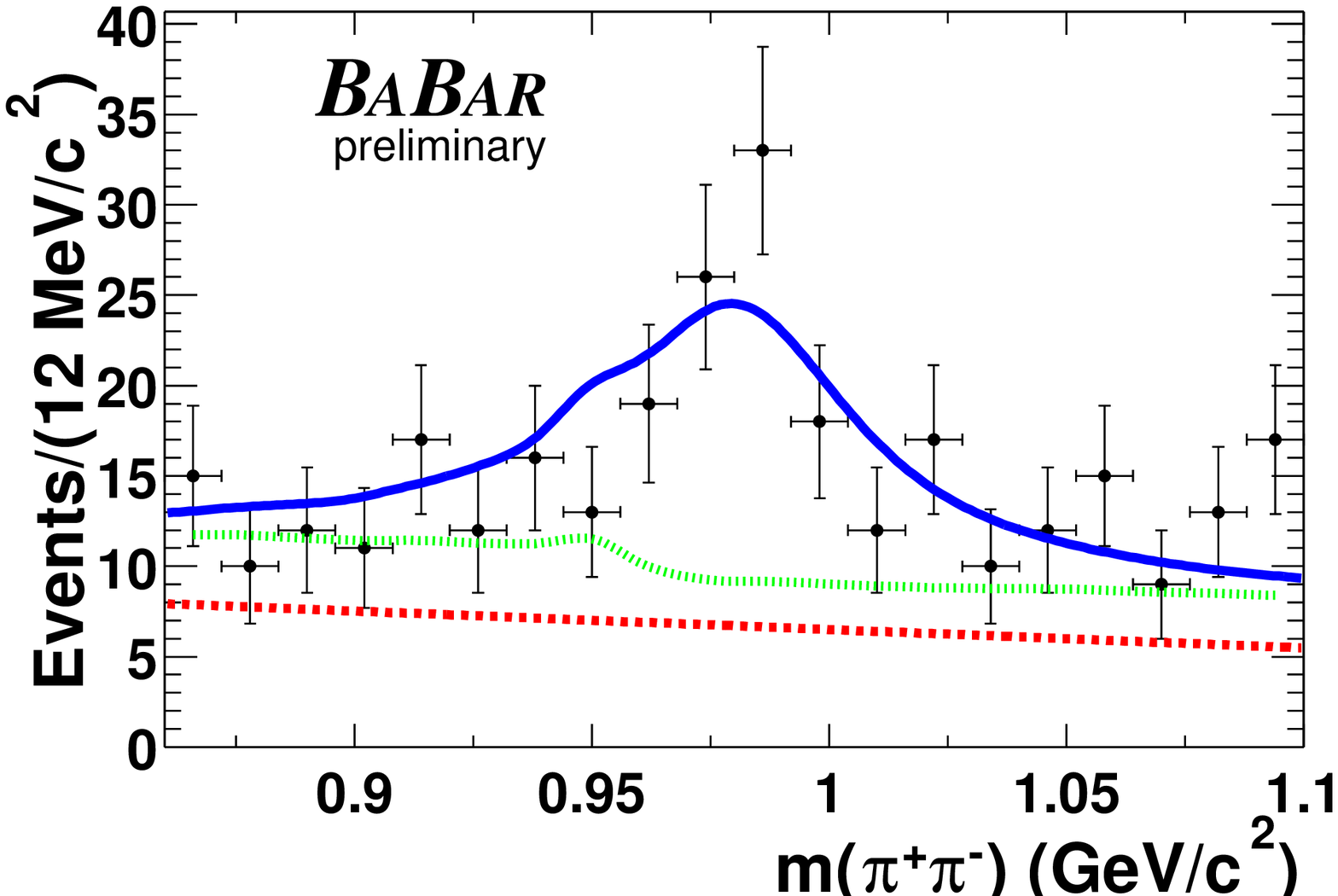}
                \epsfysize5.0cm\epsffile{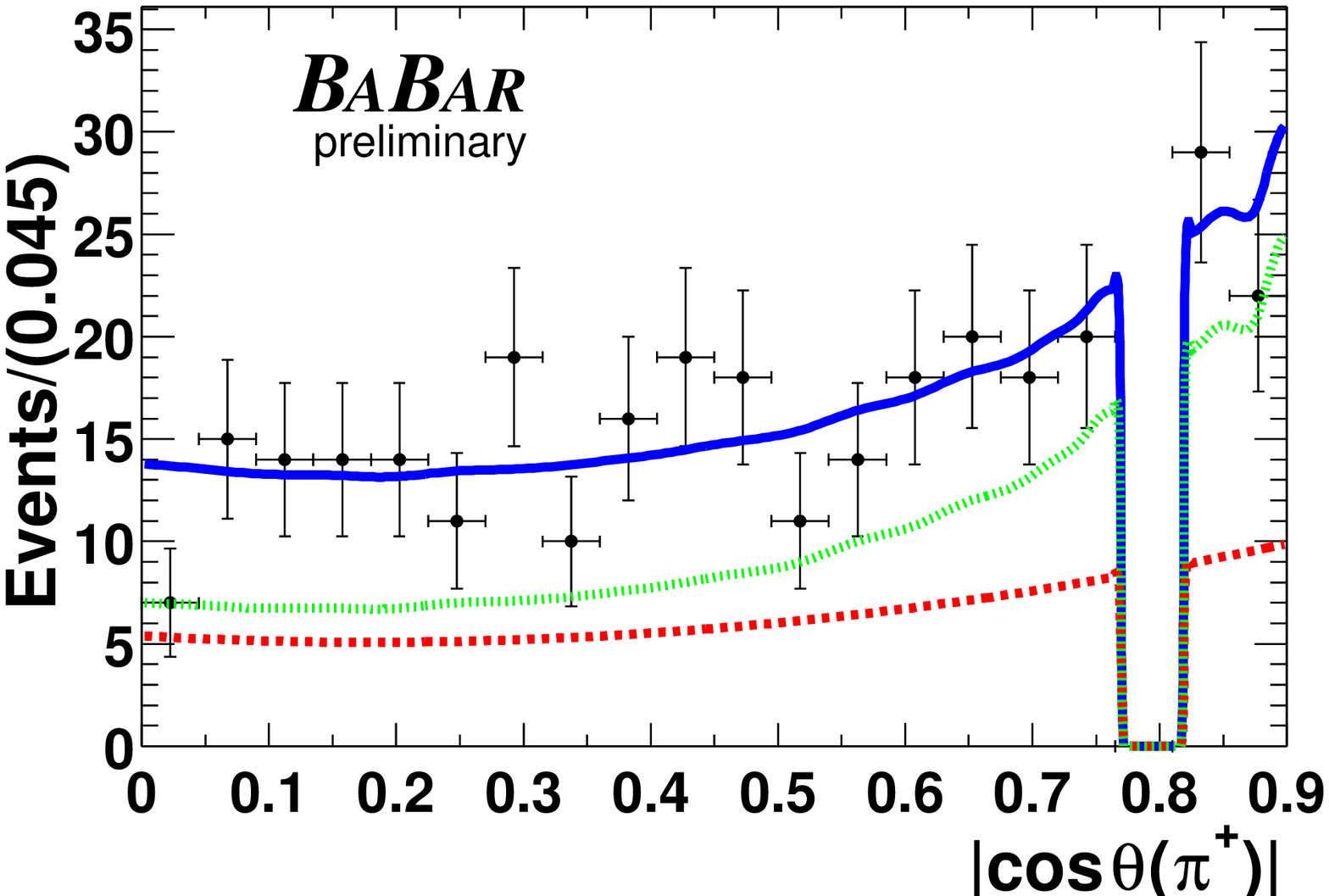}} 
\vspace{0.45cm}
  \centerline{ 	\epsfysize5.0cm\epsffile{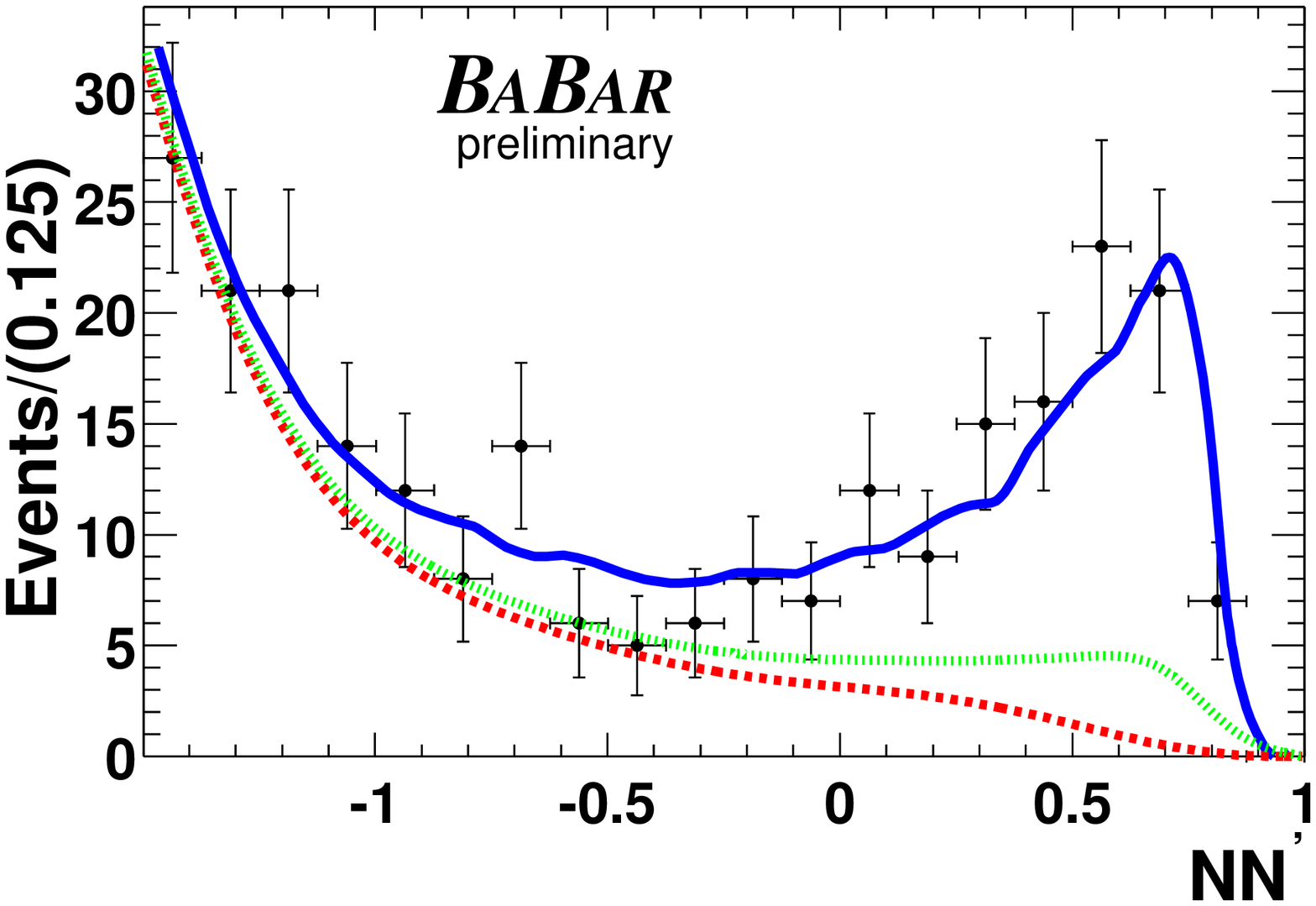}}
\caption{\label{fig:ProjMesDE}
	Distributions of  (clockwise from top left) $\de$, $\mes$, 
	$|\cos{\theta(\pi^+)}|$, $m(\pi^+\pi^-)$ and the NN output for 
  samples enhanced in $\fzeros\KS$ signal (purity is $\sim 45\%$.)	
  The solid curve represents 
  a projection of the maximum-likelihood fit result. The dashed 
	curve represents the contribution from continuum events, and 
	the dotted line (middle) indicates the combined contributions from 
	continuum events and $B$ backgrounds.  For presentation purposes, the region  
	$0.765<|\cos\theta(\pi^+)|<0.81$ has been removed to suppress 
	the contribution from $D^-\pi^+$ events.}
\end{figure}

\section{Fit Results}
\label{sec:results}

The maximum-likelihood fit results in the \CP-violation parameters:
\begin{eqnarray*}
	S	& = & -0.95^{+0.32}_{-0.23} \pm 0.10\, ,\\
	C	& = & -0.24\pm 0.31         \pm 0.15 \,,
\end{eqnarray*}
where the first errors are statistical and the second are systematic.
The improvement in the error with respect to the previous
result ($127\times10^6$ $\FourS \to B\Bbar$ decays,
$\sigma_\textrm{stat}(S) = ^{+0.56}_{-0.51}$) is due mainly 
to the increased luminosity, but is due also to one event
with large signal probability and to the proximity of the measured
$S$ and the physical limit ($|S| \le 1$).
We find an $\fzeros\KS$ event yield of $152.4\pm18.5$ which is 
consistent with the previously measured branching fraction~\cite{f0ksprl}.

\begin{figure}[p]
  \vspace{-0.4cm}
  \hspace{-0.2cm}
  \centerline{\epsfysize16.0cm \epsffile{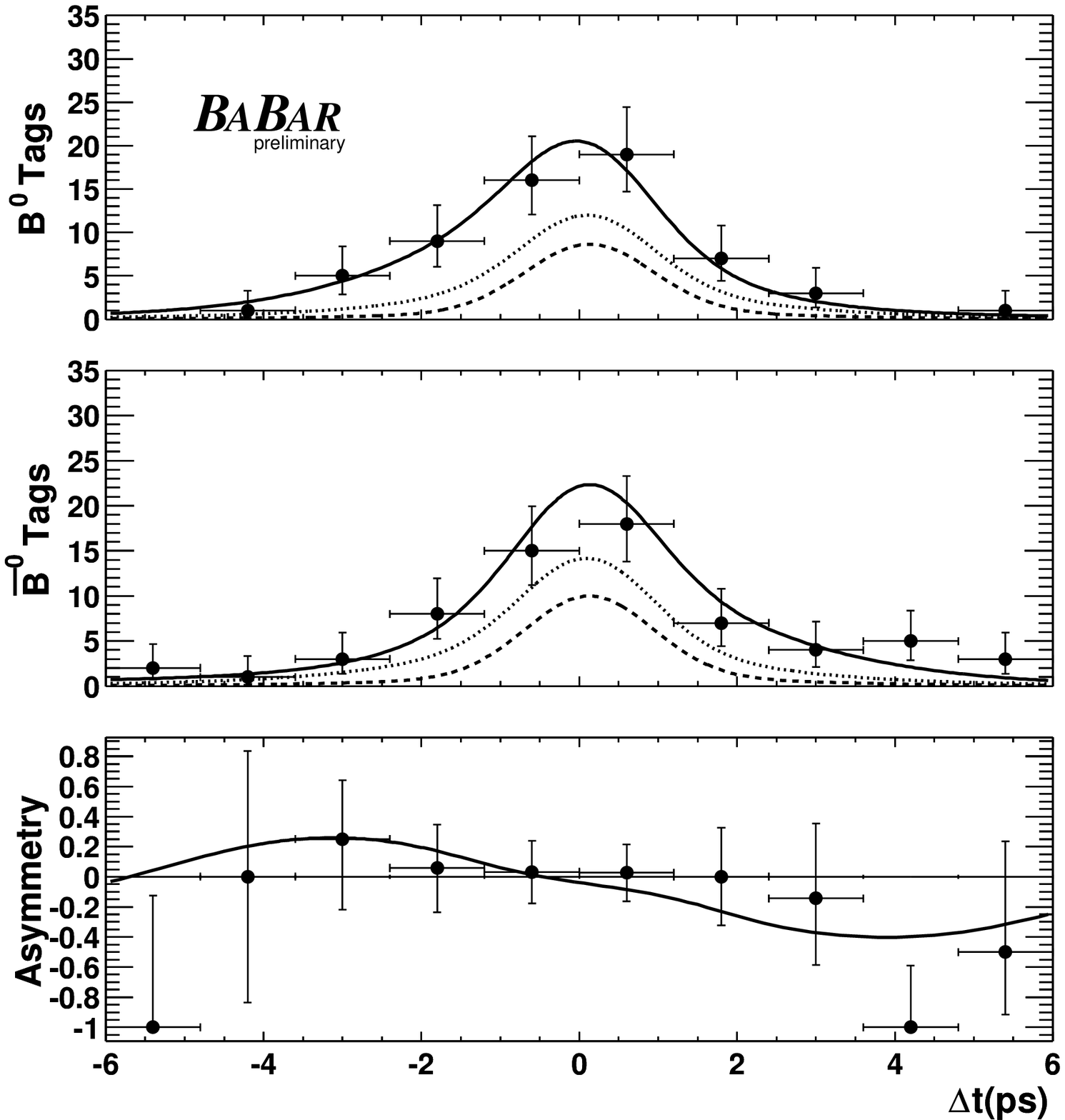}}
\vspace{-0.3cm}
\caption{
  The signal enhanced time distributions tagged as
	$\Bz_{\rm tag}$ (top) and  $\Bzb_{\rm tag}$ (middle), and the 
	asymmetry,  $A_{\Bz/\Bzb}$ (bottom). The solid curve
	is a projection of the fit result.
	The dashed line is the distribution for continuum background and the 
	dotted line is the total \B- and continuum-background 
	contribution. }
	\label{fig:asymCS} 
\vspace{-0.2in}
\end{figure}

Figure~\ref{fig:ProjMesDE} shows distributions 
of $\de$, $\mes$, $|\cos{\theta(\pi^+)}|$ and $m(\pi^+\pi^-)$,
that are enhanced in signal content by cuts on the signal-to-continuum 
likelihood ratios of the other discriminating variables. 
The  time-dependent distributions and  asymmetry
$A_{\Bz/\Bzb}=  (N_{\Bz} - N_{\Bzb})/(N_{\Bz} + N_{\Bzb})$ 
in the tagged events are presented in Fig.~\ref{fig:asymCS}.

We validated the stability of the nominal fit by testing
different fit configurations where each configuration had
a discriminating variable removed. As another cross-check, we allow
the $\Bz$ lifetime, $\tau_{\Bz}$, to vary. We find 
$\tau_{\Bz} = (1.52\pm 0.22)\ps$, in agreement with the world average
$\tau_{\Bz} = (1.536 \pm 0.014)\ps$~\cite{HFAG}, and
the remaining free parameters are consistent with the nominal fit.

\section{Summary}
\label{sec:summary}

In summary, we have presented an updated preliminary measurement of the 
\CP-violating asymmetries in $\Bz\to\ffzeroppKs$ decays. 
Our results for $S$ and $C$ are consistent with the Standard Model.
The hypothesis of no mixing-induced \CP violation is
excluded at the $2.3\sigma$ level.


\newpage

\section{Acknowledgments}
\label{sec:Acknowledgments}
We are grateful for the 
extraordinary contributions of our \pep2\ colleagues in
achieving the excellent luminosity and machine conditions
that have made this work possible.
The success of this project also relies critically on the 
expertise and dedication of the computing organizations that 
support \babar.
The collaborating institutions wish to thank 
SLAC for its support and the kind hospitality extended to them. 
This work is supported by the
US Department of Energy
and National Science Foundation, the
Natural Sciences and Engineering Research Council (Canada),
Institute of High Energy Physics (China), the
Commissariat \`a l'Energie Atomique and
Institut National de Physique Nucl\'eaire et de Physique des Particules
(France), the
Bundesministerium f\"ur Bildung und Forschung and
Deutsche Forschungsgemeinschaft
(Germany), the
Istituto Nazionale di Fisica Nucleare (Italy),
the Foundation for Fundamental Research on Matter (The Netherlands),
the Research Council of Norway, the
Ministry of Science and Technology of the Russian Federation, and the
Particle Physics and Astronomy Research Council (United Kingdom). 
Individuals have received support from 
CONACyT (Mexico),
the A. P. Sloan Foundation, 
the Research Corporation,
and the Alexander von Humboldt Foundation.


\end{document}